\documentclass[aps,pra,reprint,superscriptaddress,floatfix,longbibliography]{revtex4-1}

\usepackage{amsmath}    
\usepackage{amssymb}
\usepackage{graphicx}   
\usepackage{color}      
\usepackage{hyperref}   
\usepackage{dcolumn}

\def\lsco{La$_{2-x}$Sr$_x$CuO$_4$}

\def\ybco{YBa$_2$Cu$_3$O$_{6+x}$}

\def\bscco{Bi$_2$Sr$_2$CaCu$_2$O$_{8+\delta}$}
\def\hbco{HgBa$_2$CuO$_{4+\delta}$}
\def\ncco{Nd$_{2-x}$Ce$_x$CuO$_4$}

\begin{document}

\newcommand{\LSNO}{La$_{2-x}$Sr$_{x}$NiO$_{4}$}
\newcommand{\LSNOn}{La$_{1.75}$Sr$_{0.25}$NiO$_{4}$}

\title{Low-energy antiferromagnetic spin fluctuations limit the coherent superconducting gap in cuprates}

\author{Yangmu Li}
\affiliation{Condensed Matter Physics and Materials Science Division, Brookhaven National Laboratory, Upton, New York 11973, USA}
\author{Ruidan Zhong}
\thanks{Present address: Department of Chemistry, Princeton University,
Princeton, New Jersey 08544, USA}
\affiliation{Condensed Matter Physics and Materials Science Division, Brookhaven National Laboratory, Upton, New York 11973, USA}
\author{M. B. Stone}
\author{A. I. Kolesnikov}
\affiliation{Neutron Scattering Division, Oak Ridge National Laboratory, Oak Ridge, Tennessee 37831, USA}
\author{G. D. Gu}
\author{I. A. Zaliznyak}
\author{J. M. Tranquada}
\email{jtran@bnl.gov}
\affiliation{Condensed Matter Physics and Materials Science Division, Brookhaven National Laboratory, Upton, New York 11973, USA}

\date{\today} 

\begin{abstract}
Motivated by recent attention to a potential antiferromagnetic quantum critical point at $x_c\sim 0.19$, we have used inelastic neutron scattering to investigate the low-energy spin excitations in crystals of La$_{2-x}$Sr$_x$CuO$_4$ bracketing $x_c$.   We observe a peak in the normal-state spin-fluctuation weight at $\sim20$~meV for both $x=0.21$ and 0.17, inconsistent with quantum critical behavior.  The presence of the peak raises the question of whether low-energy spin fluctuations limit the onset of superconducting order.  Empirically evaluating the spin gap $\Delta_{\rm spin}$ in the superconducting state, we find that $\Delta_{\rm spin}$ is equal to the coherent superconducting gap $\Delta_c$ determined by electronic spectroscopies.  To test whether this is a general result for other cuprate families, we have checked through the literature and find that $\Delta_c\le\Delta_{\rm spin}$ for cuprates with uniform $d$-wave superconductivity.  We discuss the implications of this result.
\end{abstract}

\maketitle

\section{Introduction}

The layered copper-oxide compounds continue to be intriguing because they combine exotic ordered states, such as high-temperature superconductivity, with a challenging electronic environment that defies simple description \cite{keim15}.  The parent compounds are charge-transfer Mott insulators, for which charge excitations face a gap of $\sim2$~eV \cite{baso05}, while the only low-energy excitations are the antiferromagnetic spin waves associated with the magnetic moments localized on Cu \cite{kast98}.  By chemical substitution or addition of oxygens in the spacer layers, it is possible to introduce holes into the CuO$_2$ planes.  The holes would like to delocalize to minimize their kinetic energy, but this competes with the local superexchange interactions, resulting in complex inhomogeneous correlations \cite{oren00,lee06}, with a significant modification of the spin excitations from the antiferromagnetic state \cite{fuji12a}.  Another consequence is that, at low concentrations, only the doped holes contribute to the low-temperature transport properties;  with increasing doping, the effective carrier density begins to rise faster than the dopant density \cite{padi05,gork06}.  It rises quite rapidly as the doped-hole density $p$ approaches a putative pseudogap critical point at $p_c \sim 0.19$ \cite{tall01,bado16}.

\begin{figure}[t]
 \centering
    \includegraphics[width=0.9\columnwidth]{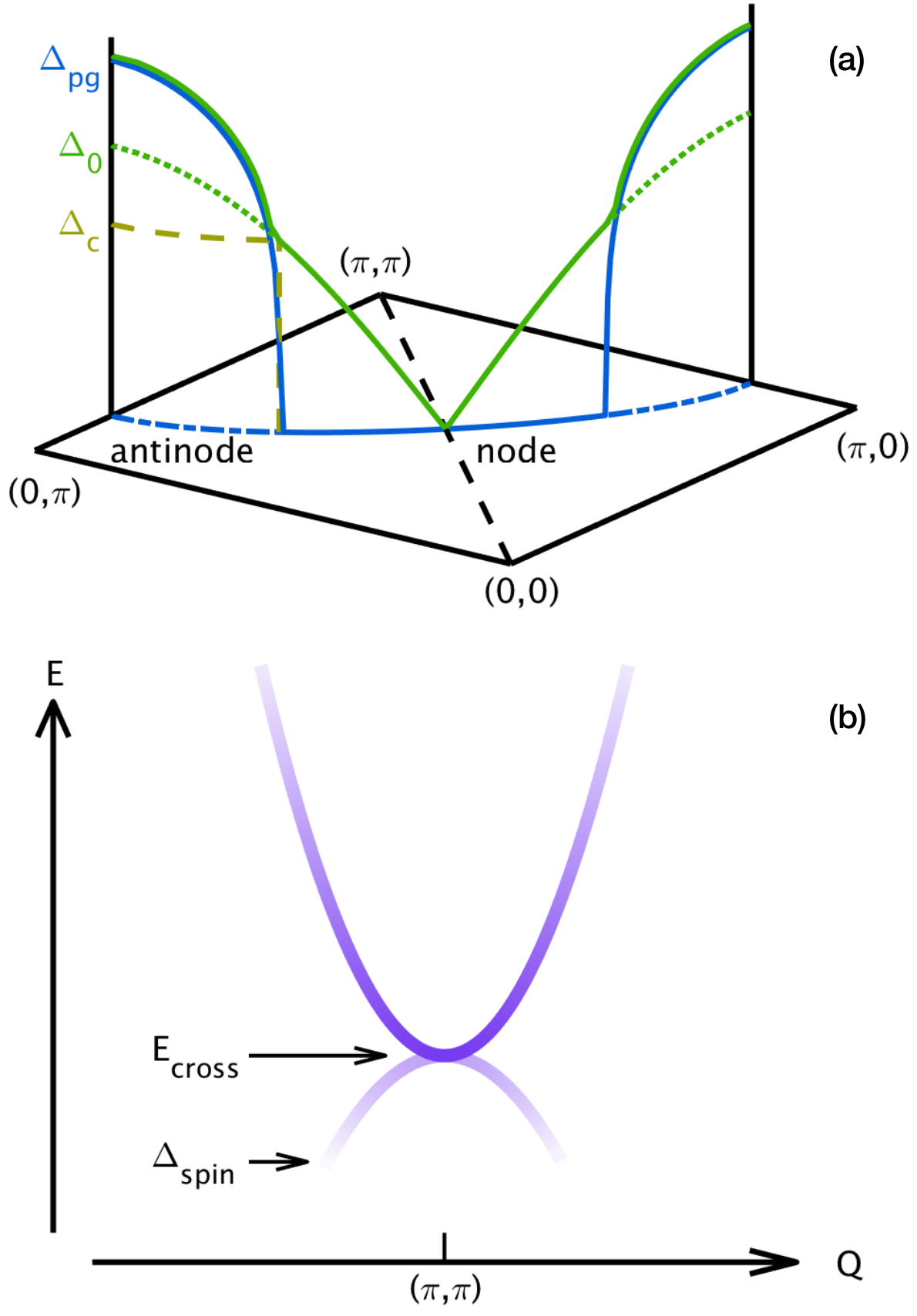}
    \caption{\label{fg:schem}  (a) Schematic of electronic gaps around a quadrant of reciprocal space for a square CuO$_2$ plane with lattice constant $a=1$.  Green line: below $T_c$, a $d$-wave gap is found near the node, extrapolating to $\Delta_0$, but switches to the pseudogap in the antinodal region, rising to $\Delta_{\rm pg}$.  Blue line: above $T_c$, coherent gap (with maximum energy of $\Delta_c$) closes to form a Fermi arc. (b) Schematic spin excitation spectrum as a function momentum transfer {\bf Q}, defining the spin gap $\Delta_{\rm spin}$ and $E_{\rm cross}$. }
\end{figure}

Given the prominence of antiferromagnetic spin fluctuations, it is commonly (but not universally) believed that they play a role in the hole pairing that is essential for superconductivity \cite{lee06,kive07,scal12a}.  It has also motivated proposals that the pseudogap critical point might be associated with an antiferromagnetic quantum critical point \cite{bado16,eber16}.   We set out to test this possibility by using inelastic neutron scattering to measure the spin fluctuations in the system \lsco\  (LSCO) at doping levels $p=x=0.17$ and 0.21, superconducting compositions close to but bracketing $p_c$.  If critical fluctuations are important, then we might expect the normal state to exhibit spin fluctuations spread over a substantial energy range with no characteristic energy scale.  Instead, we found an effective normal-state spin gap of similar magnitude in both samples.  Furthermore, the spin gap $\Delta_{\rm spin}$ identified by the shift in magnetic spectral weight on cooling below the superconducting transition temperature $T_c$ is approximately equal to the coherent superconducting gap $\Delta_c$  that has been determined previously by measurements such as Andreev reflection \cite{deut99}, and electronic Raman scattering \cite{deve07}; it corresponds to the magnitude of the $d$-wave gap at the wave vectors at the ends of the normal-state Fermi arcs (delimited by the pseudogap) as determined by angle-resolved photoemission spectroscopy (ARPES) \cite{lee07,kami15} and scanning tunneling spectroscopy (STS) \cite{push09,kuro10,fuji14a}, and indicated in Fig.~\ref{fg:schem}(a).

Is this behavior unique to LSCO?  To check, we looked through the literature to identify neutron scattering measurements of $\Delta_{\rm spin}$ and corresponding Raman scattering results for $\Delta_c$.  We find that, for all cuprate families studied, these data satisfy the relation $\Delta_c \le \Delta_{\rm spin}$ in the regime where uniform $d$-wave superconductivity occurs.   It is intriguing that the correlation is with $\Delta_c$ rather than $2\Delta_c$, as one might have expected from models that assume proximity to a Fermi-liquid state \cite{esch06}.  In fact, we argue that the gap relationship indicates the incompatibility of Bogoliubov quasiparticles of a spatially-uniform $d$-wave superconductor with the antiferromagnetic spin excitations of local Cu moments \cite{ande97a}.  

The rest of the paper is organized as follows.  We describe the experimental methods in the next section.  The analysis of our data, as well as a detailed comparison with results from the literature, is presented in Sec.~\ref{sc:anal}.  This is followed by a discussion of the results and a comparison with theoretical analyses.  We conclude with a brief summary.

\begin{figure}[t]
 \centering
    \includegraphics[width=0.9\columnwidth]{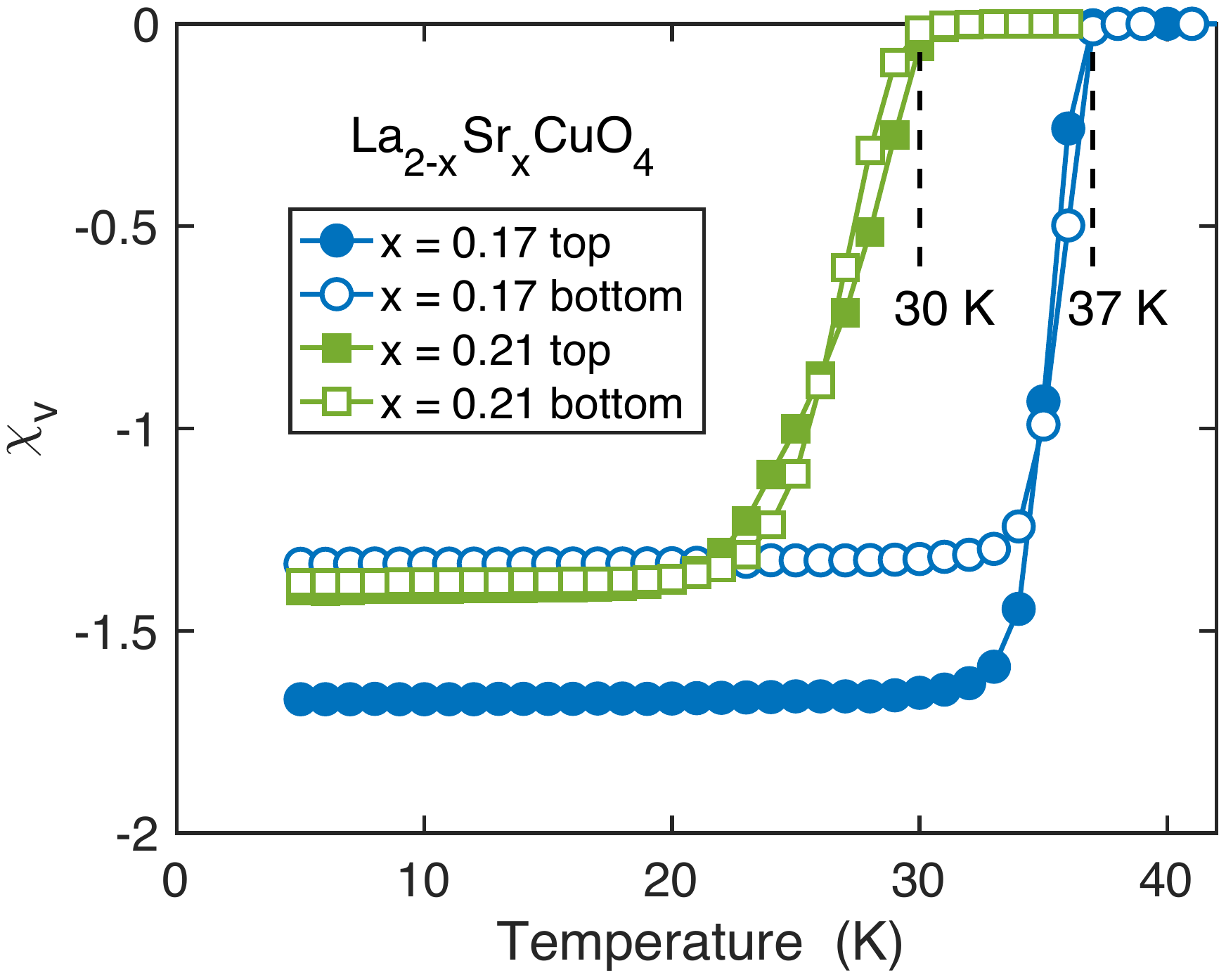}
    \caption{\label{fg:susc}   Volume susceptibility measured for samples of LSCO with $x=0.17$ and 0.21.  For $x=0.17$, bottom (top) was 4.5 cm (15 cm) from the start of growth; for $x=0.21$, bottom (top) was 6 cm (14 cm) from the start of growth. }
\end{figure}

\section{Experimental Methods}

Cylindrical crystals of LSCO 7-mm in diameter were grown for $x=0.17$ and 0.21 by the traveling-solvent floating-zone method.  
For each composition, a single feed rod of length 20--25 cm was used; after growth, the initial several centimeters of the crystal rod were removed and discarded, while the remainder was annealed in flowing O$_2$ at 980 $^\circ$C for 1 week.  The superconducting transition temperatures, 37 and 30 K, respectively, were determined by dc magnetization measurements with a field of 10 G applied (after cooling in zero field).  Figure \ref{fg:susc} shows the measurements for small pieces (0.1--0.3 g) taken from the bottom and the top of each crystal rod.  (The orientation of crystal axes and crystal shape with respect to the field was arbitrary; differences in orientation contribute to the apparent variation in magnitude of the diamagnetic response.  No correction was made for shape anisotropy, which is likely responsible for the susceptibility exceeding the full-shielding response of $\chi_v=-1$.)  The variation of $T_c$ across each sample rod is less than 1~K.  

For the $x=0.17$ sample, three crystals, with a total mass of $\sim35$~g, were coaligned by x-ray Laue diffraction with a tetragonal $[110]$ axis in the vertical direction.  For the $x=0.21$ sample, four crystals, with a total mass of 25.5~g, were coaligned with a tetragonal $[100]$ in the vertical direction using the neutron alignment station CG-1B at the High Flux Isotope Reactor, Oak Ridge National Lab (ORNL).  To describe the scattering data, we will use a unit cell based on the low-temperature orthorhombic phase with $a\approx b\approx 5.35$~\AA, $c\approx 13.2$~\AA\ \cite{rada94}, and wave vectors ${\bf Q} = (h,k,l)$ in units of $(2\pi/a,2\pi/b,2\pi/c)$.  The $x=0.17$ crystals were co-aligned with [100] and [001] axes in the horizontal plane; for $x=0.21$, [110] and [001] were in the plane.

The inelastic neutron scattering measurements were performed at the SEQUOIA time-of-flight spectrometer (BL-17) of the Spallation Neutron Source, ORNL \cite{sequoia10}.  For each composition, the sample was mounted in a closed-cycle helium refrigerator for temperature control.  Measurements were performed with an incident energy of either 30 or 60 meV, using the the high-resolution or high-flux chopper, respectively, operating at a frequency of 420 Hz, for sample temperatures of base  (4 or 5~K), 36~K (near $T_c$), and 300~K.  For each condition, data were collected with the in-plane sample orientation relative to the incident beam rotated in $1^\circ$ steps over a range of at least $120^\circ$.  Initial data reduction was performed with {\tt MANTID} \cite{mantid14}.  Reference measurements on a vanadium standard were used to convert the intensity data to absolute units, and we extracted the imaginary part of the dynamical susceptibility, $\chi''({\bf Q},\omega)$, in the conventional fashion \cite{xu13}, making use of the magnetic form factor for Cu$^{2+}$ hybridized with 4 in-plane O neighbors \cite{walt09}.

In choosing to measure at 36~K, we intended to be as close as possible to $T_c$ for $x=0.17$, so that we would not be excessively above $T_c$ for $x=0.21$.  Working without the susceptibility data in hand, we chose a measurement temperature that is just slightly below $T_c$ for $x=0.17$; however, given the finite width of the transition seen in Fig.~\ref{fg:susc}, 36~K is at the start of the transition, and we are confident that had the measurements been performed at 37~K, the results would be virtually the same.

\section{Analysis}
\label{sc:anal}
\subsection{Data reduction}

\begin{figure}[t]
 \centering
    \includegraphics[width=1.0\columnwidth]{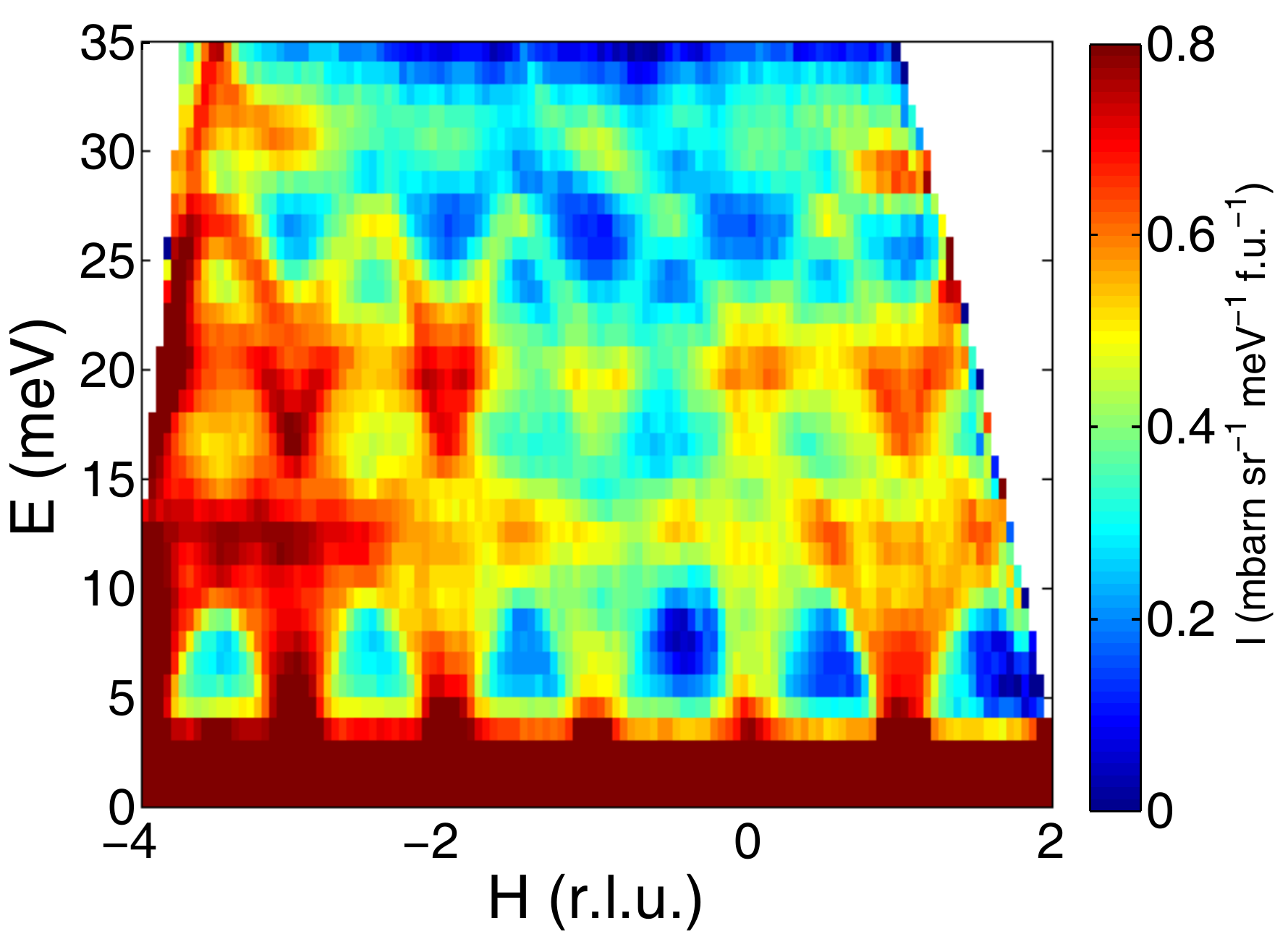}
    \caption{\label{fg:phonons}  Scattered intensity as function of energy and ${\bf Q}=(H,0,0)$ for LSCO $x=0.17$ at 36~K.  The intensity has been integrated over $-0.2\le K\le0.2$ and over all measured $L$ within the range $-2\le L\le 5$. Antiferromagnetic scattering is allowed around $H=1,\ -1,\ -3$.  The range of sample orientations measured was selected so that, for $H=-1$, the $L$ range sampled was within $-2\le L\le2$; the $L$ ranges at other $H$ values vary considerably.  The difference in phonon intensity at $H = -1$ and 1 is due to different sampled $L$ ranges.}
\end{figure}

Although phonon scattering is relatively weak at small $Q$, so is the magnetic scattering.  Figure~\ref{fg:phonons} shows a plot of scattered intensity along ${\bf Q}=(H,0,0)$; the phonon contributions are obviously quite significant even at $H=-1$, where the $L$ range has been optimized to emphasize the magnetic response.  The neutron scattering cross section is proportional to
\begin{equation}
  S({\bf Q},\omega) \sim \chi''({\bf Q},\omega)/(1-e^{-kT/\hbar\omega}),
\end{equation}
where $S({\bf Q},\omega)$ is the dynamic structure factor and $\chi''({\bf Q},\omega)$ is the imaginary part of the dynamic susceptibility.  Previous work has shown that the magnetic contribution to $\chi''$, at least for $\hbar\omega \lesssim20$~meV, decreases with increasing temperature \cite{aepp97}, whereas the phonon part is essentially constant with temperature.  At 300~K, the magnetic contribution to $\chi''$ is weak relative to the phonon part; hence, we will subtract the measured $\chi''$ at 300~K from the low temperature data to largely remove the phonon contribution.

\begin{figure}[t]
 \centering
    \includegraphics[width=1.0\columnwidth]{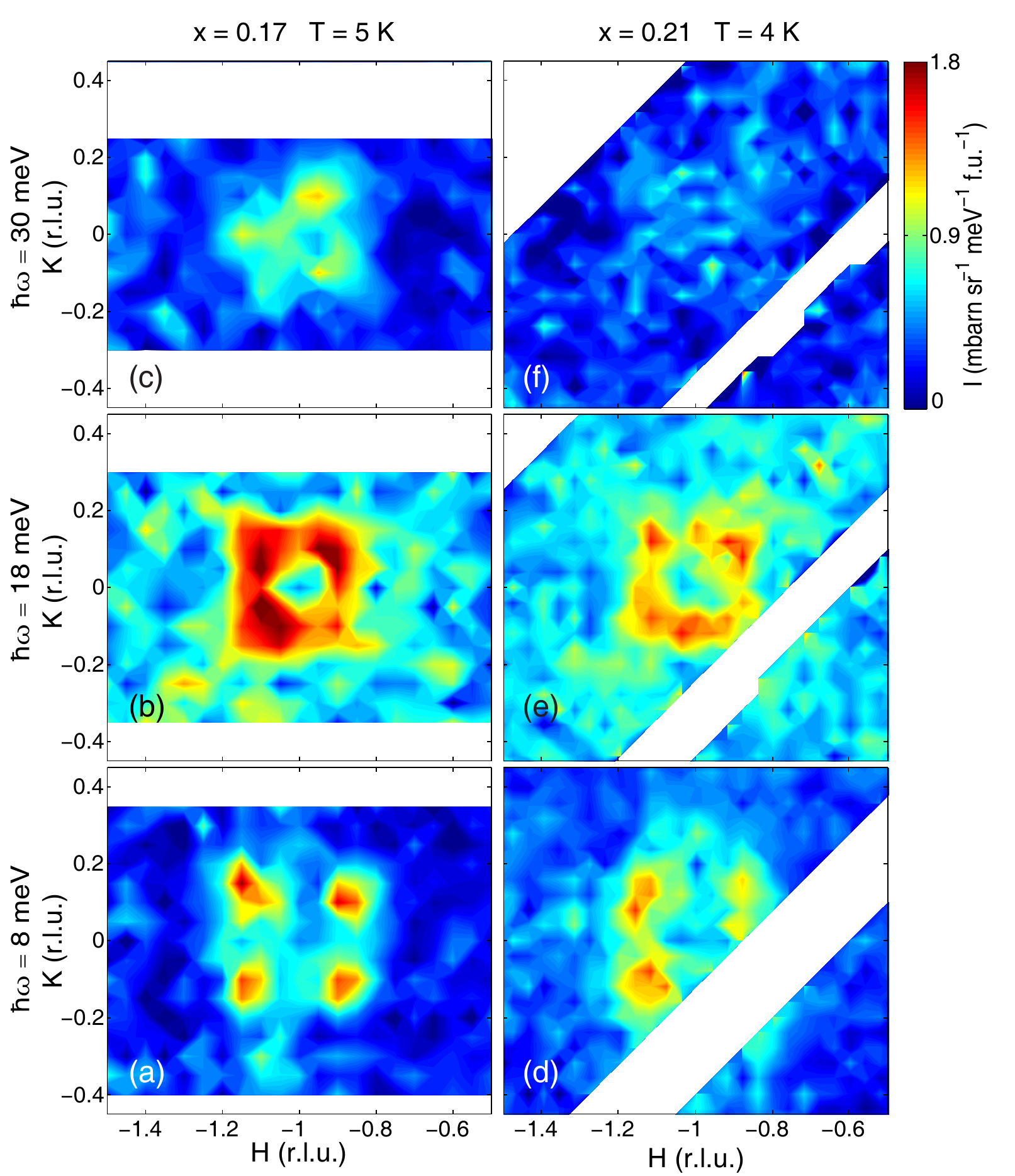}
    \caption{\label{fg:constE}   Constant-energy slices of magnetic scattering intensity (after subtraction of phonon contribution) as a function of in-plane wave vector for the $x=0.17$ (left) and 0.21 (right) samples at base temperature.  Excitation energies: 8 / 18 / 30~meV (bottom / middle / top); energy width of 1~meV; incident energy of 60 meV.  White areas correspond to detector gaps.}
\end{figure}

\begin{figure}[t]
 \centering
    \includegraphics[width=1.0\columnwidth]{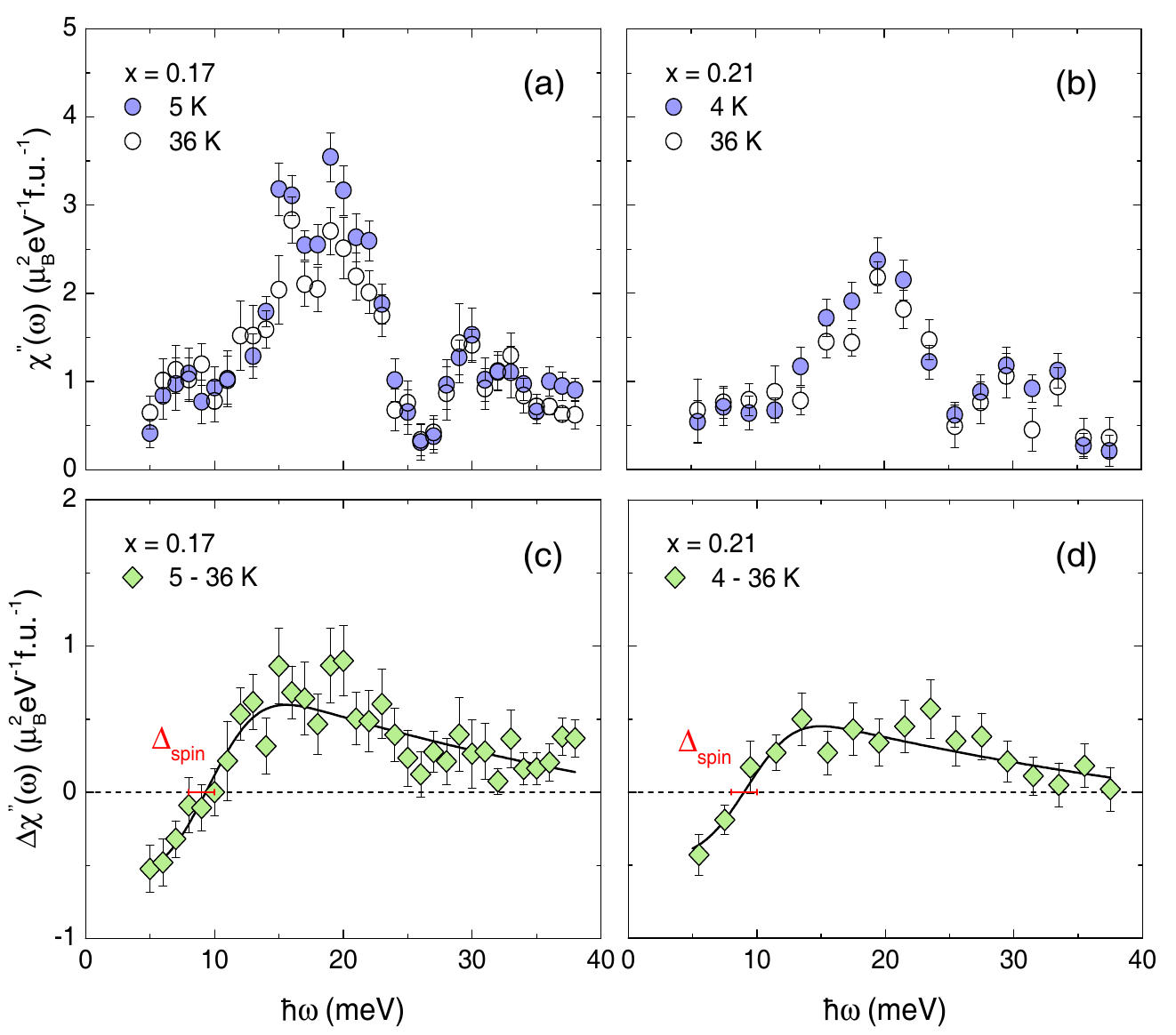}
    \caption{\label{fg:diff}  {\bf Q}-integrated $\chi''(\omega)$ for (a) $x=0.17$ and (b) $x=0.21$ at base temperature and $T=36$~K; the energy bin width is twice as large for the $x=0.21$ results to compensate for the different sample orientation with respect to the spectrometer.  Difference in $\chi''(\omega)$ between base temperature and 36~K for (c) $x=0.17$ and (d) $x=0.21$.  (The differences were taken using data without phonon correction, as the phonon contribution cancels almost completely.) In the latter, the experimental $\Delta_{\rm spin}$, discussed in the text, is indicated in red. }
\end{figure}

Figure~\ref{fg:constE} shows constant-energy slices of the low-temperature magnetic $\chi''$ at excitation energies of 8, 18, and 30~meV as a function of the in-plane wave vector, where the antiferromagnetic wave vector corresponds to ${\bf Q}_{\rm AF} = (1,0,0)$.   (We have integrated over momentum transfer perpendicular to the plane, covering $-2<L<5$.) The usual incommensurate peaks are seen at low energy, consistent with previous studies \cite{vign07,lips07}.   By averaging the magnetic response over the first Brillouin zone, we obtain the {\bf Q}-integrated response $\chi''(\omega)$.  Figure~\ref{fg:diff}(a) and (b) show the results for $\chi''({\bf Q})$ at $T=36~{\rm K} \sim T_c$ and at $T\ll T_c$.  For each sample, there is a peak at $\sim 20$~meV, though the response is clearly weaker for the $x=0.21$ sample.  

The underlying cause of this peak in the magnetic spectrum has not been fully established, but a coupling between phonons and superexchange has been proposed \cite{wagm15,wagm16}.  The key observation here is that the peak is still clearly resolved in the normal state of a sample with $p>p_c$.  The presence of a clear energy scale is inconsistent with expectations for critical magnetic fluctuations.  Hence, our results provide evidence against a generic antiferromagnetic quantum critical point in cuprates.  Instead, evidence points to a crossover with $p$ from stronger to weaker correlations \cite{fuji12a,huss18}.

\begin{figure}[t]
 \centering
    \includegraphics[width=1.0\columnwidth]{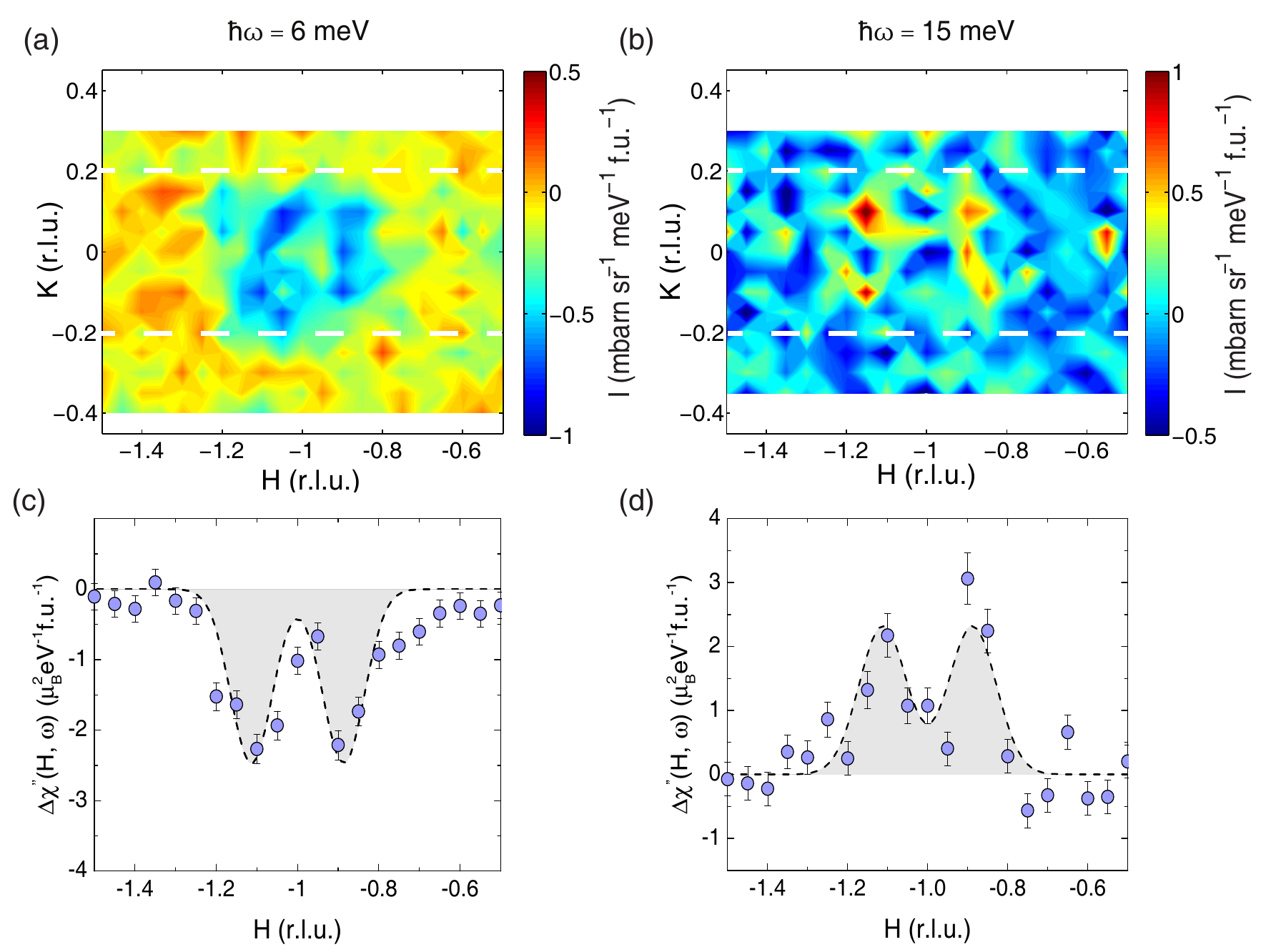}
    \caption{\label{fg:sub}  Analysis of $\Delta\chi''=\chi''(5\ K)-\chi''(36\ K)$ for LSCO $x=0.17$.  Constant-energy slices of $\Delta\chi''$ for (a) $\hbar\omega= 6$~ meV, (b) 15~meV.  (c) and (d) are the corresponding cuts obtained after integrating over $K$ within the range denoted by the dashed white lines in (a) and (b); a constant background difference has been subtracted.  The dashed lines in (c) and (d) are fitted gaussian peaks used to evaluate the magnetic contribution to $\Delta\chi''(\omega)$. }
\end{figure}

While subtracting the high-temperature phonon contribution does a reasonably good job of isolating the magnetic response, the differences between $\chi''(\omega)$ below and near $T_c$, as shown in Fig.~\ref{fg:diff}(a) and (b), are small.  To determine the change in $\chi''$ across $T_c$ while minimizing the uncertainty, we take the difference in the un-corrected $\chi''$, in which case the phonon signal cancels out directly, and fit the results.  Examples of such differences for the $x=0.17$ sample are shown in Fig.~\ref{fg:sub}.  [Note that the background should be temperature independent, but converting the measured intensity to $\chi''$ scales the background in a temperature-dependent fashion.  To account for this, we fit a constant background term when evaluating $\Delta\chi''$; this background shift has already been subtracted in Fig.~\ref{fg:sub}(c) and (d).]

\subsection{Identifying the spin gap}

The change in $\chi''(\omega)$ between base temperature and 36~K is shown in Fig.~\ref{fg:diff}(c) and (d), demonstrating the development of a spin gap in the superconducting state and a shift of weight from below to above the gap.  To be explicit, we choose to define the spin gap $\Delta_{\rm spin}$ to be the energy at which the temperature difference in $\chi''(\omega)$ crosses zero, which corresponds to $9\pm1$ meV for both samples.  

Our definition of the spin gap is different from what has been used previously.  In two early studies of LSCO $x=0.15$, the gap was defined as the energy below which the magnetic signal is essentially zero, yielding a gap of 3.5~meV \cite{yama95a,peti97}.  In another pair of studies on LSCO $x=0.14$ \cite{maso92} and $x=0.163$ \cite{lake99}, the measured $\chi''$ was fit with formulas having parallels with electronic spectroscopy, yielding gaps of 6~meV and 6.7~meV, respectively.  Given the current state of theory for the cuprates, we are not aware of any generally accepted definition of the spin gap.  We believe that our choice of definition is appropriate for comparison with results taken from electronic spectroscopies (see below).  Applying our definition to the data in \cite{lake99} yields $\Delta_{\rm spin} \approx 9$~meV.

Comparing with $\chi''(\omega)$ at 36 K, we note that the amount of spectral weight that shifts through the superconducting transition is small (consistent with previous observations of limited weight in the ``resonance'' peak \cite{kee02}), and $\Delta_{\rm spin}$ occurs in a range where the spectral weight in the normal state is already weak.  This raises the question as to whether the energy cost of gapping low-energy spin fluctuations limits the coherent superconducting gap $\Delta_c$.

\subsection{Comparing with the coherent superconducting gap}

To test this idea, we need to compare $\Delta_{\rm spin}$ with measurements of $\Delta_c$.  While ARPES studies have provided evidence for a coherent gap scale associated with the normal-state Fermi arc \cite{lee07,kami15}, Raman scattering data are available for a larger variety of cuprates.  In particular, Raman spectra with $B_{2g}$ symmetry probe electronic states near the $d$-wave gap node, and measurements below $T_c$ yield an intensity peak at $\Omega(B_{2g})=2\Delta_c$.  Raman studies on LSCO find that $\Omega(B_{2g})$ is essentially independent of doping for a significant range of $x$, and corresponds to $\Delta_c\approx8$~meV for $x=0.15$ and 0.17 \cite{musc10,suga13}.  (We will assume an uncertainty of $\pm15$\%\ in the values of $\Delta_c$ estimated from Raman results, as the peak at $\Omega_{B_{2g}}$ tends to have significant width.)  Andreev reflection data suggest that this trend extends out to $x=0.2$ \cite{gonn01}.  Hence, we find that $\Delta_c \approx \Delta_{\rm spin}$, within the experimental uncertainties, for our LSCO samples.

This result suggests that low-energy ungapped spin fluctuations may limit the coherent superconducting gap.  If this is true for LSCO, it ought to be true for other cuprates, as well.  We evaluate the latter next.

\begin{table}[b]
\caption{\label{tab:data} Results for the spin gap and $E_{\rm cross}$ for a variety of cuprates; $p/n$ corresponds to an estimate of the doped hole/electron concentration.  Values in parentheses are uncertainties in meV.}
\begin{ruledtabular}
\begin{tabular}{cccccc}
Compound & $p$/$n$  & $T_c$ & $\Delta_{\rm spin}$ & $E_{\rm cross}$ & Refs. \\
        & & (K) & (meV) & (meV) &  \\
\hline
\ybco & 0.10 & 59 & 15(5) & \ \ 32.5 & \cite{stoc05} \\
 & 0.11 & \ \ 62.7 & 20(5) & 34 & \cite{hayd04} \\
 & 0.16 & 93 & 28(5) & 41 & \cite{rezn08} \\
 & 0.17 & \ \ 92.5 & 27(5) & 41 & \cite{woo06} \\
\bscco & 0.16 & 93 & 32(3) & 43 & \cite{fong99} \\
  & 0.18 & 91 & 30(5) & 40 & \cite{xu09} \\
  & 0.19 & 87 & 32(5) & 42 & \cite{fauq07} \\
  & 0.20 & 83 & 30(4) & 38 & \cite{he01} \\
  & 0.21 & 70 & 24(5) & 34 & \cite{capo07} \\
\hbco & 0.10 & 71 & 28(5) & 50 & \cite{chan16b} \\
 & 0.13 & 88 & 40(5) & 59 & \cite{chan16c} \\
\lsco & 0.16 & \ \ 38.5 & \ \ 8(1) & 45 & \cite{lake99,vign07} \\
 & 0.17 & 37 & \ \ 9(1) & 45 & this work \\
 & 0.21 & 30 & \ \ 9(1) &     & this work \\
\ncco & 0.15 & 25 & \ \ \ \ 4.5(1) & & \cite{yama03,zhao07c} \\
\end{tabular}
\end{ruledtabular}
\end{table}

\subsection{Testing the relationship on other cuprates}

We have gone through the literature and identified studies of various cuprates that allow a reasonable estimate for $\Delta_{\rm spin}$.  The results are listed in Table~\ref{tab:data}, where we have also reported values for $E_{\rm cross}$, as defined in Fig.~\ref{fg:schem}(b).  We note that, in all cases, $\chi''(\omega)$ is weak in the normal state at and below $\hbar\omega=\Delta_{\rm spin}$.  

Raman scattering studies on \ybco, \bscco, and \hbco\ have found that $2\Delta_c\approx 6kT_c$ \cite{munn11,guya08}, which allows us to estimate $\Delta_c$ from $T_c$.  For \hbco, we chose to interpolate the results of Li {\it et al.} \cite{li13}, while for Nd$_{1.85}$Ce$_{0.15}$CuO$_4$, Blumberg {\it et al.} \cite{blum02b} found $2\Delta_c=4.4kT_c$.

\begin{figure}[t]
 \centering
    \includegraphics[width=0.7\columnwidth]{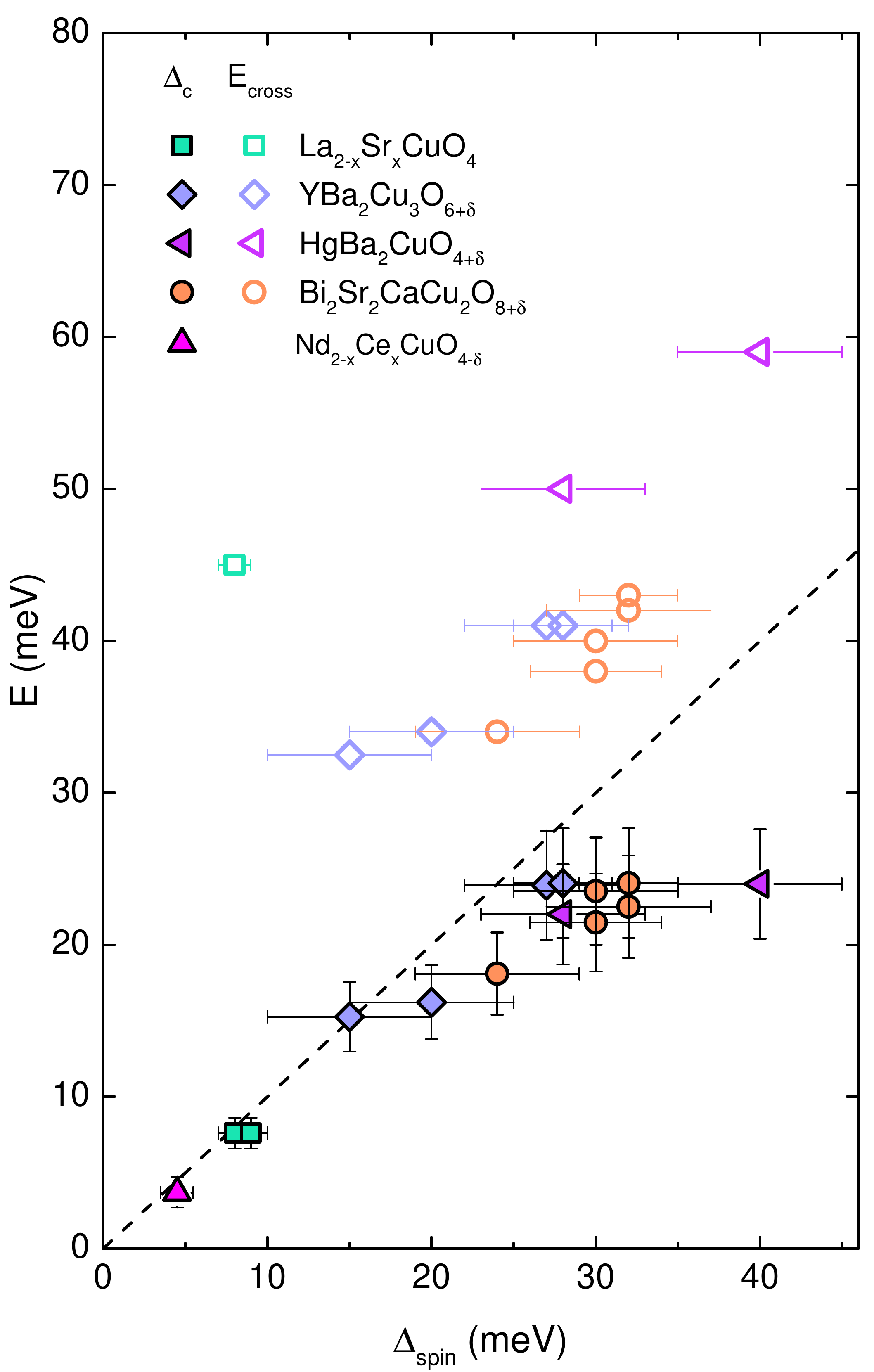} 
    \caption{\label{fg:gaps} Plot of $\Delta_c$ and $E_{\rm cross}$ vs.\ $\Delta_{\rm spin}$, corresponding to the data and references in Table I. }
\end{figure}

We plot $\Delta_c$ and $E_{\rm cross}$ vs.\ $\Delta_{\rm spin}$ in Fig.~\ref{fg:gaps}.  We find that $\Delta_c\le\Delta_{\rm spin}$ for all studied cuprates, whereas $E_{\rm cross}$ does not show a simple universal relationship with either $\Delta_{\rm spin}$ or $\Delta_c$.  (The fly in the ointment is LSCO, where $E_{\rm cross}$ is comparable to that of other cuprates but $\Delta_c$ and $\Delta_{\rm spin}$ are much smaller.)  Where there is a spin gap present in the normal state that is already larger than $\Delta_c$, as in the case of underdoped \hbco\ \cite{chan16b,chan16c}, then the spin fluctuations do not limit the development of superconducting coherence and there is no significant shift in magnetic spectral weight across $T_c$.

We note that analysis of in-plane tunneling measurements on \ybco\ with $T_c\approx 90$~K yield $\Delta_c = 28(3)$~meV \cite{wei98}.  While this value of $\Delta_c$ is slightly larger than the value of 23~meV estimated from $T_c$, it is still compatible with $\Delta_c\le \Delta_{\rm spin}$ given the $\Delta_{\rm spin}$ values listed in Table~\ref{tab:data} and taking uncertainties into account.

\section{Discussion}

Our focus on the spin gap and its relation to the coherent superconducting gap differs from more-commonly discussed relationships, so we will try to put it in perspective.  Many early theoretical analyses of magnetic excitations in metallic cuprates started from a weak coupling approach, analyzing the spin response in terms of the excitation of free electrons across the Fermi surface, with interactions treated in the Random Phase Approximation \cite{ruva92,lu92,bulu93,si93}.  When the superconducting gap opens, the interactions can pull spin excitations below $2\Delta_0$ (the lower bound for spin excitations of noninteracting electrons), resulting in a spin resonance peak \cite{liu95,onuf02,norm07,esch06}.  Such a model appeared to give a reasonable description of initial experimental results on the enhancement of commensurate magnetic excitations below $T_c$ in \ybco\ \cite{ross92,mook93,fong96} and \bscco\ \cite{fong99,he01}.  Of course, these early studies only probed commensurate scattering near $E_{\rm cross}$; later studies established the dispersion away from $E_{\rm cross}$, and the presence of that dispersion in the normal state \cite{arai99,bour00,hayd04,tran04,stoc05,hink07,xu09,chan16b}.  Furthermore, it has been demonstrated that the dispersion evolves continuously from the antiferromagnetic-insulator phase, where one has local Cu moments coupled by superexchange \cite{dai01,birg06,fuji12a}.  Recent numerical studies of dynamical correlations in Hubbard models with onsite Coulomb repulsion comparable to the bandwidth support this perspective \cite{huan17,noce18}. Thus, there is good reason to believe that the antiferromagnetic excitations detected by neutron scattering across the phase diagram are predominantly from local Cu moments coupled by superexchange \cite{huck08}.  

As noted by Anderson \cite{ande97a}, the superexchange interaction does not involve quasiparticles.  Hence, the interaction of quasiparticles with the spin correlations is inevitably strong. 
Non-Fermi-liquid behavior can result from charge carriers scattering off of spin fluctuations \cite{vekh04}.   The impact of such interactions is evident in the transport properties of cuprates.  For example, deviations from $\rho\sim T$ in \ybco\ correspond to the onset of the gap in antiferromagnetic spin fluctuations observed by nuclear magnetic resonance \cite{ito93,taki91,baek12}.

Millis, Sachdev, and Varma \cite{mill88} showed for a $d$-wave superconductor that spin fluctuations with energies below a critical threshold are pair breaking.  Dahm and Scalapino \cite{dahm18} recently analyzed the dependence of the superconducting gap and transition temperature on $\chi''({\bf Q},\omega)$ for underdoped \ybco, using experimentally-determined spin-fluctuation data \cite{dahm09}.  They found that the spin excitations that disperse upwards from $E_{\rm cross}$ enhance $T_c$, while those below $E_{\rm cross}$ tend to be bad for superconductivity.

Those theoretical results are consistent with the empirical evidence that the presence of low-energy spin fluctuations correlates with a reduced $T_c$; however, they do not directly address the relationship between $\Delta_{\rm spin}$ and $\Delta_c$.  To understand that, we note that when a spin gap is present, quasiparticles with energies below the gap can propagate without interacting with the spins, leaving them sharply-defined in the near-nodal regime, as observed by ARPES.  Towards the antinodal region, quasiparticles with energies above $\Delta_{\rm spin}$ can decay by creating spin fluctuations.  That contribution to the electronic self-energy can make it unfavorable for those electronic states to participate in the coherent superconducting state.  This perspective is consistent with the idea that the pseudogap detected by ARPES is a consequence of AF correlations \cite{zaki17}.  Furthermore, all of the cuprates for which quantum oscillations have been observed in high magnetic fields \cite{tail09,seba15,bari13} have large spin gaps \cite{huck14}, consistent with sharp quasiparticles at energies close to the chemical potential.  In \ybco, the effective mass diverges as $p$ is reduced towards 0.085 \cite{seba12}, which correlates with the closing of the spin gap \cite{dai01,hink08,baek12}.

\begin{figure}[t]
 \centering
    \includegraphics[width=0.9\columnwidth]{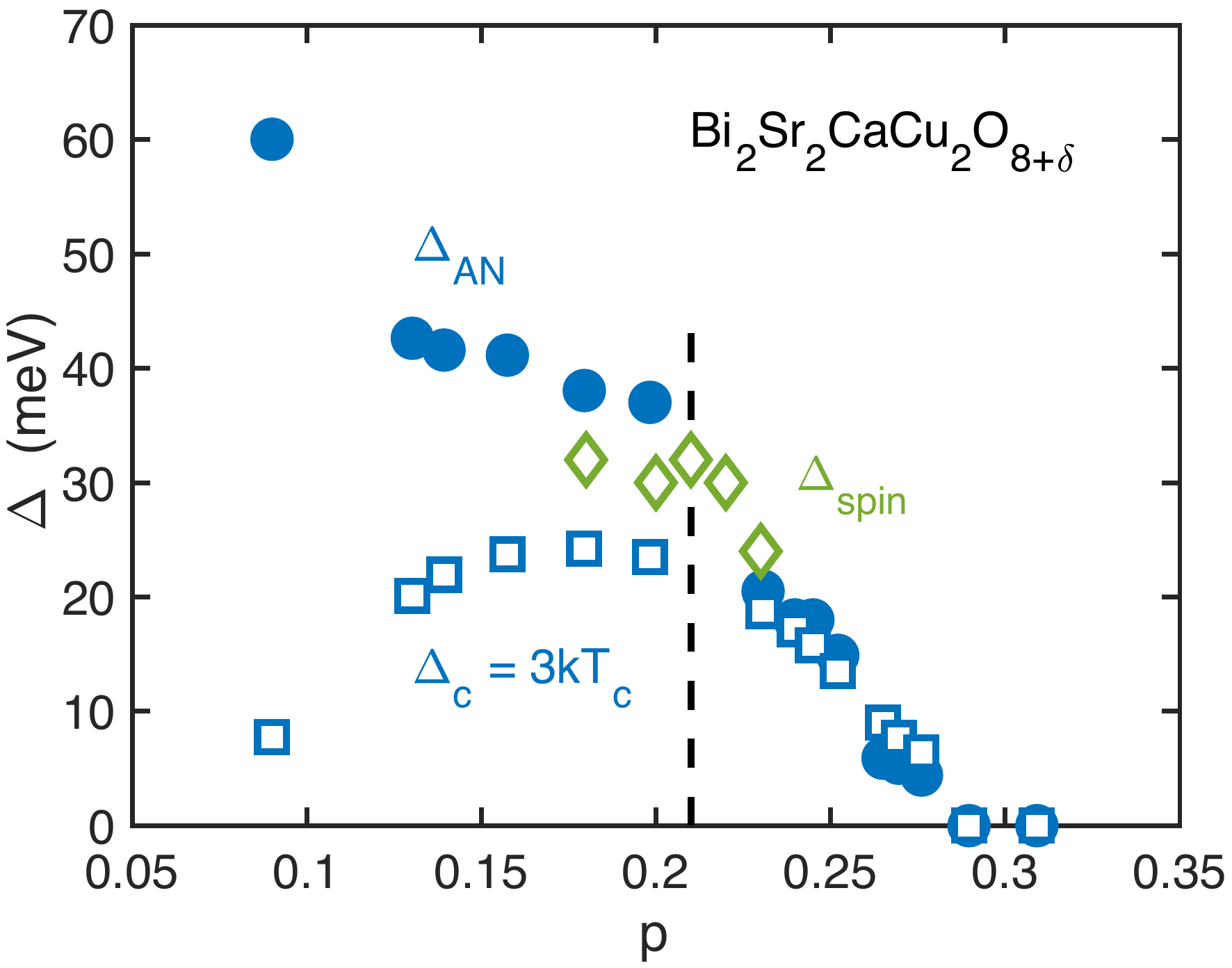} 
    \caption{\label{fg:ARPES} Comparison of the antinodal quasiparticle energy at $T<T_c$ in \bscco\  \cite{droz18} (filled circles), $\Delta_c$ estimated from $T_c$ (open squares), and the reported values of $\Delta_{\rm spin}$ from Table~\ref{tab:data} (open diamonds).  Note that we have used the values of $p$ reported in \cite{droz18}, which were determined from the size of the Fermi surface, resulting in an increase in $p$ by 0.02 from typical estimates based on $T_c(p)$ \cite{ober92}.  To put the spin gap results on the same scale, we have increased the corresponding values of $p$ listed in Table~\ref{tab:data}  by 0.02.  The vertical dashed line indicates the adjusted $p$ value at which an STS study \cite{fuji14a} found a crossover in the Bogoliubov quasiparticle distribution from a finite arc to the full Fermi surface.}
\end{figure}
 
Recent ARPES results covering a broad range of doping in \bscco\ (Bi2212) appear to be quite consistent with our analysis \cite{he18,zhon18b,droz18}.  For under- and optimally-doped samples, the quasiparticle peak near the antinodal (AN) wave vector is broad and has an energy of 30 meV or greater.  As the doping increases beyond $p_c$ and  $T_c$ decreases below 70 K,  the AN peak suddenly sharpens and its energy drops below 20~meV.   From Table~\ref{tab:data}, we see that $\Delta_{\rm spin}\approx 24$~meV for overdoped Bi2212 with $T_c=70$~K. To illustrate this, we have plotted in Fig.~\ref{fg:ARPES} the AN peak energy from \cite{droz18} as a function of $p$, together with $\Delta_c$ estimated from $T_c$ and the results for $\Delta_{\rm spin}$ listed in Table~\ref{tab:data}.  It appears that the superconducting gap becomes coherent around the entire Fermi surface when $\Delta_0$ drops below $\Delta_{\rm spin}$.  This interpretation is consistent with the observation of a crossover of Bogoliubov quasiparticles from an arc to a full Fermi surface at $p_c$ by STS \cite{fuji14a}.  (We note that Ref.~\cite{he18} interprets the change in AN electronic self-energy in terms of electron-phonon coupling.  While we do not dispute that e-ph coupling may play some role, we argue that the dominant effect is associated with spin fluctuations.)

Evidence for strong damping of quasiparticles interacting with spin correlations also comes from measurements on samples  with no spin gap, such as LSCO with $x<0.13$ \cite{chan08,jaco15}.  For example,  an ARPES study on LSCO $x=0.08$ found a broad spectral function at the node (compared to $x=0.145$) with a peak well below the Fermi energy even for $T<T_c$ \cite{razz13}.  This feature was interpreted as a nodal superconducting gap.  We suggest that this is a signature of strong damping due to the presence of gapless AF spin fluctuations.  Indeed, Loder {\it et al.} \cite{lode11} found a gap in Hartree-Fock calculations of a pair-density-wave state, including spin-stripe order. Similar behavior is also seen in Bi2212 in the most-underdoped, yet superconducting, samples \cite{vish12}.  The absence of coherent quasiparticles in underdoped LSCO is also evident from the large magnitude of the in-plane resistivity \cite{ando01,emer95b} and the fact that the low-temperature resistivity shows insulating behavior when the superconductivity is suppressed by a strong magnetic field \cite{ando95}.

Of course, while sharply-defined quasiparticles are the starting point for the BCS model \cite{bard57}, they are not a general prerequisite for coherent superconductivity \cite{pate18}.  In cuprates, there is reason to believe that this occurs by Josephson coupling between local regions with strong hole pairing, just as has been proposed for the most extreme case of stripe ordered La$_{1.875}$Ba$_{0.125}$CuO$_4$ \cite{li07,berg07,frad15,emer97}.  While it may seem surprising that superconductivity could occur without normal-state quasiparticles, we note that this is exactly what happens along the $c$ axis in most underdoped cuprates, where transport is incoherent and Josephson coupling is essential \cite{katt08}.

\section{Summary}

Our neutron scattering investigation of low-energy spin fluctuations in LSCO near $x_c\sim0.19$ has led to two significant conclusions.  First, the presence of structure in the energy dependence of magnetic spectral weight in the normal state is inconsistent with antiferromagnetic critical behavior.  Second, looking at the shift in the magnetic spectral weight across $T_c$ leads to the proposal that low-energy spin fluctuations limit the superconducting order.  Comparisons with the coherent superconducting gap defined by electronic spectroscopies leads to the relation $\Delta_c\le \Delta_{\rm spin}$.  This relation is closely connected to the crossover from superconducting coherence on a finite arc ($\Delta_c < \Delta_0$) for $p<p_c$ to coherence all around the nominal Fermi surface ($\Delta_c=\Delta_0$) for $p>p_c$.  

\medskip
\section{Acknowledgments}

We thank V. Fanelli for valuable assistance at SEQUOIA, and S. Kivelson, A. Chubukov, and A. Tsvelik for constructive comments.  Work at Brookhaven is supported by the Office of Basic Energy Sciences, Materials Sciences and Engineering Division, U.S. Department of Energy (DOE) under Contract No.\ DE-SC0012704.   A portion of this research used resources at the Spallation Neutron Source and the High Flux Isotope Reactor, DOE Office of Science User Facilities operated by Oak Ridge National Laboratory.

\bibliography{LNO,theory,neutrons,misc}

\end{document}